\documentclass[conference]{IEEEtran}
\usepackage{cite}
\usepackage{graphicx}

\usepackage{amsmath, amssymb, amsfonts}
\usepackage{caption, enumerate, enumitem}
\usepackage{epsfig, epstopdf}
\usepackage{adjustbox}
\usepackage{ragged2e}
\usepackage{float}
\usepackage{bm}
\usepackage{xcolor}
\usepackage{graphicx}
\usepackage{makecell}
\usepackage[top=1in, bottom=1.2in, left=0.7in, right= 0.7in]{geometry}

\usepackage{tabularray}
\UseTblrLibrary{amsmath}

\usepackage{stfloats}
\usepackage{balance}

\newtheorem{definition}{Definition}
\newtheorem{lemma}{Lemma}
\newtheorem{theorem}{Theorem}
\newtheorem{example}{Example}

\newtheorem{remark}{Remark}

\newtheorem{note}{Note}

\newlist{mycases}{enumerate}{1}
\setlist[mycases]{label=\textit{Case~\Roman*:},align=left,itemindent=0pt,leftmargin=0pt,labelwidth=-4pt}

\newlist{mysubcases}{enumerate}{1}
\setlist[mysubcases]{label=\textit{Sub-case~\arabic*:},align=left,itemindent=0pt,leftmargin=0.6cm,labelwidth=-4pt}

\def\BibTeX{{\rm B\kern-.05em{\sc i\kern-.025em b}\kern-.08em
		T\kern-.1667em\lower.7ex\hbox{E}\kern-.125emX}}

\begin{document}
	
	\title{Systematic Construction of Golay Complementary Sets of Arbitrary Lengths and Alphabet Sizes\\
	}
	
	\author{\IEEEauthorblockN{Abhishek Roy}
		\IEEEauthorblockA{\textit{Department of Mathematics}\\
			\textit{Indian Institute of Technology Patna}\\
			\textit{Bihar, India- 801106}\\
			Email: \texttt{abhishekroy@ieee.org}}
		\and
		\IEEEauthorblockN{Sudhan Majhi}
		\IEEEauthorblockA{\textit{Department of ECE} \\
			\textit{Indian Institute of Science, Bangalore}\\
			Karnataka, India- 560012\\
			Email: \texttt{smajhi@iisc.ac.in}}
		\and
		\IEEEauthorblockN{Subhabrata Paul}
		\IEEEauthorblockA{\textit{Department of Mathematics} \\
			\textit{Indian Institute of Technology Patna}\\
			Bihar, India- 801106 \\
			Email: \texttt{subhabrata@iitp.ac.in}}
	}
	
	\maketitle
	
	\begin{abstract}
		One of the important applications of Golay complementary sets (GCSs) is the reduction of peak-to-mean envelope power ratio (PMEPR) in orthogonal frequency division multiplexing (OFDM) systems. OFDM has played a major role in modern wireless systems such as long-term-evolution (LTE), 5th generation (5G) wireless standards, etc. This paper searches for systematic constructions of GCSs of arbitrary lengths and alphabet sizes. The proposed constructions are based on extended Boolean functions (EBFs). For the first time, we can generate codes of independent parameter choices.
	\end{abstract}
	
	
	\section{Introduction}
	The importance of alphabet size has recently come to the forefront of the sequence design research community \cite{palashQCSS}. The foundation and theoretical development on the research of signaling alphabets have been done long ago in 1952 \cite{alphabet}. The authors found that the Shannon capacity (C), which is given by 
	$$C= B \log_2 \left(1+ \frac{S}{N} \right),$$
	where $B$ is the bandwidth of the channel, $S$ is the signal power, and $N$ is the noise power, cannot be achieved without using a flexible choice of alphabets in the data bits. For binary channels, the channel capacity is given by 
	$$C=1+ p \log_2(p) + (1-p) \log_2 (1-p),$$
	where $p$ is the probability of error of a data bit. To elaborate more, the authors showed that for an alphabet with $p < 1/4$, one can achieve a rate higher than $R_{0,b}- \epsilon$, where
	$$R_{0,b}=1+ 2p \log_2(2p) + (1-2p) \log_2 (1-2p),$$
	with a suitable choice of alphabet. The analogous argument also holds true for non-binary channels. The corresponding value of $R_{0,b}$ for non-binary channel is given by 
	$$R_{0,nb}=\log_2(n+1)+ 2p \log_2 \left(\frac{2p}{n} \right) + (1-2p) \log_2 (1-2p),$$
	where $(n+1)$ is the number of digits used to construct a letter of the alphabet, and the probability of error of each digit is set to be $\frac{p}{n}$.
	In fact, considering an additive white Gaussian noise (AWGN)-channel \cite{shannon}, the authors in \cite{alphabet} have explicitly constructed different alphabets, which produces good efficiency plot of information-rate vs. signal-to-noise ratio (SNR) or probability of error. 
	
	There are two notions of alphabets in information theory. The one we have already discussed is the set of data symbols to be transmitted via the wireless channel through proper encoding. The second notion that fits in the sequence design aspect is related to the construction of sequences for spread-spectrum communications. For spreading sequences, the number of digits, which was $(n+1)$ for the first case, is known as alphabet size, and we distinguish it with the notation $q$ ($q>1$ is a natural number). In spread spectrum systems, such as multi-carrier code division multiple access (MC-CDMA), the data is modulated according to the spreading sequence before transmission. The alphabet size $q$ plays the role of making the sequence elements uni-modular, explicitly saying, $q$-th roots of unity. The use of spreading sequences started with the famous Golay complementary pair (GCP) \cite{Golay51, golay1961}, which shows the ideal aperiodic auto-correlation function (AACF) sum property. Ever since its discovery, it has been widely used in modern telecommunication systems, as the AACF property is suitable for interference-free signal detection in radar \cite{radar1, radar2}, reduction of peak-to-mean envelope power ratio (PMEPR) in orthogonal frequency division multiplexing (OFDM) system \cite{davis}, etc. But, as there are practical limits on how many such Golay sequences can be obtained, Tseng and Liu proposed the concept of the Golay complementary set (GCS), which holds the same ideal AACF sum property but with more than $2$ sequences \cite{GCStseng}.
	
	Till date, there have been many constructions of GCSs of variable lengths and alphabet sizes, both with direct methods, i.e., function-based, and indirect methods, i.e., initial sequence-based. The direct methods can be found in \cite{paterson, schmidt, ChenCCC2008, ChenGCS-TIT, ChenGCS2017,  ChenGCS2018, palashCS2021, ChenMOGCS2021, palashCCC, ShenCCC}, and the indirect ones in \cite{parkerGCS, WuGCS2008, MAgcs, ZWangGCS2, WuGCS, ChenGCS2020, ShenGCS, AvikGCS2019, AvikGCS, ZWangCCC2021, LwangGCS, ZWangCCC2023, du2024golay}. The problem with these direct constructions is that they are mostly even-phased, except \cite{palashCCC, ShenCCC}. Although \cite{palashCCC, ShenCCC} offers more flexibility in the alphabet size for the sequence elements, it depends on the GCS length. Considering these limitations, in this paper, we propose a function-based direct construction of GCSs with the length and alphabet size chosen arbitrarily and independently. 
	
	The rest of the paper is organized as follows: Section II describes the useful definitions and tools for the construction. The main construction is proposed in Section III. The conclusion is drawn in Section IV.
	
	\section{Preliminary}
	In this section, we define some preliminary notions and the definitions for the subsequent sections and subsections.
	\begin{definition}[ACCF]
		For a complex-valued sequence pair, $\left( \mathbf{a}, \mathbf{b}\right)$, where $\mathbf{a}=\left( a_1, a_2, \dots, a_L\right)$, $\mathbf{b}=\left( b_1, b_2, \dots, b_L\right)$, the aperiodic cross-correlation function (ACCF) at time shift $\tau$ is defined as
		\begin{equation}
			\rho(\mathbf{a}, \mathbf{b})(\tau)=
			\begin{cases}
				\sum_{i=1}^{L-\tau} a_{i} b_{i+\tau}^*, & 0 \leq \tau <L;\\
				\sum_{i=1}^{L+\tau} a_{i-\tau} b_{i}^*, & -L < \tau <0;\\
				0, & \text{otherwise},
			\end{cases}
		\end{equation}
		where $(\cdot)^*$ denotes the complex conjugate.
	\end{definition}
	
	\begin{note}
		When $\mathbf{a}=\mathbf{b}$, then the function is called the AACF and denoted by $\rho(\mathbf{a})(\tau)$.
	\end{note}
	
	\begin{definition}[GCS]
		A GCS is a set of sequences $\mathbf{C}= \{\mathbf{a}_1, \mathbf{a}_2, \dots, \mathbf{a}_M\}$ of fixed length $L$ (or codes) that exhibit the following property:
		\begin{equation}
			\sum_{\lambda=1}^{M} \rho(\mathbf{a}_{\lambda}) (\tau)= 0, \tau \neq 0.
		\end{equation}
		In this paper, we denote a GCS of set size $M$, length $L$, and alphabet size $q$ as an $(q, M, L)$-GCS.
	\end{definition}
	
	\begin{note}
		The property of a GCS is the ideal auto-correlation property. When $M=2$, a GCS reduces to GCP.
	\end{note}
	
	\subsection{Extended Boolean Function (EBF)}
	The EBF is an extension of the generalized Boolean function (GBF). For a set of integers modulo $N$ defined by $\mathbb{Z}_N=\{0,1, \dots, N-1\}$, a GBF is defined to be any function $f: \mathbb{Z}_2^m \rightarrow \mathbb{Z}_q$ which is a $\mathbb{Z}_q$-linear combination of the monomials:
	$$\mathcal{M}=\{1,x_1, x_2, \dots, x_m, x_1 x_2, \dots, x_1 x_2 \cdots x_m\},$$
	where $2 \mid q$, and the operations are done in the modulo $q$ settings. This function form, called the algebraic normal form (ANF), is generally unique to a GBF. 
	An EBF can be regarded as a GBF-like function where we extend the domain of the GBF. Specifically, an EBF is defined as $f: \mathbb{Z}_p^m \rightarrow \mathbb{Z}_q$, which may be defined as the $\mathbb{Z}_q$-linear combination of the monomials: 
	$$\mathcal{M}'= \left\{ \prod_{\alpha=1}^{m}  x_\alpha^{j_\alpha} \middle| j=\left(j_1, j_2, \dots, j_m\right) \in  \mathbb{Z}_p^{m} \right\},$$
	where $p$ and $q$ are positive integers satisfying $p \mid q$.
	
	One can associate a $\mathbb{Z}_q$-valued and a complex-valued sequence corresponding to an EBF. Let $\Vec{i}= \left(i_1, i_2, \dots, i_m\right)$ be the $p$-ary representation of the integer $i= \sum_{l=1}^{m} i_l p^{l-1}$, where $0 \leq i \leq p^m-1$. Then the $\mathbb{Z}_q$-valued sequence $\mathrm{f}$
	corresponding to the EBF $f$ is given by
	\begin{equation}
		{\bf{f}}= \left( f(\vec{0}), f(\vec{1}), \dots, f(\overrightarrow{p^m-1})\right).
	\end{equation}
	The complex-valued sequence $\psi(f)$ is defined similarly as
	\begin{equation}
		\psi(f)= \left(\zeta_q^{f(\vec{0})}, \zeta_q^{f(\vec{1})}, \dots, \zeta_q^{f(\overrightarrow{p^m-1})}  \right),
	\end{equation}
	where $\zeta_q= e^{2\pi \sqrt{-1}/q}$ is the $q$-th primitive root of unity. Also, $\psi_L(f)$ denotes the complex-valued sequence derived from the sequence $\psi(f)$ after deleting the last $p^m-L$ elements. Similarly, $\phi_{L}(f)$ denotes the $\mathbb{Z}_q$-valued sequence derived from $\bf{f}$ after deleting the last $p^m-L$ elements.

	\subsection{Peak-to-Mean Envelope Power Ratio (PMEPR)}
	The envelope power is an essential property of a transmission system of OFDM. It, in a way, measures the power efficiency of the system. PMEPR should be ideally low for efficient transmission. For a $\mathbb{Z}_q$-valued sequence $\mathbf{a}= \left(  a_1, a_2, \dots, a_L\right)$, the PMEPR \cite{davis} is defined as
	\begin{equation}
		\text{PMEPR} \left( \mathbf{a} \right)= \sup_{0 \leq \Delta f t \leq 1} \frac{\left|  
			S_{\mathbf{a}} (t)\right| ^2}{L},
	\end{equation}
	where $	S_{\mathbf{a}} (t)= \sum_{i=1}^{L} \zeta_q^{a_i + q f_i t }$ is the complex envelope of the transmitted signal, $f_i= f+ (i-1) \Delta f$, $f$ is the carrier frequency of the first sub-carrier of OFDM, $\Delta f$ is the sub-carrier spacing in OFDM, $t$ represents the time instance.
	
	\begin{lemma}[\cite{davis}]\label{lemma_PMEPR}
		For a GCS with flock size $M$, the PMEPR of each sequence is bounded by $M$.  
	\end{lemma}

	\section{Proposed Construction}
	In this section, we propose the construction of a general $q$-ary GCS with arbitrary lengths with the help of EBFs. But, before construction, we make some remarks regarding the flexibility of the length.
	
	\begin{remark}
		For any $p \geq 1$ and $L \geq 1$, which are integers, $\exists m \in \mathbb{N}$ such that $p^{m-1} \leq L < p^m$. What it implies is that for any $p, L \geq 1$ the value of $m$ is fixed and depends on both $p, L$. With notation, this can be stated as:
		$$m= \mathcal{F}(p,L)$$
		for some function $\mathcal{F}$.
	\end{remark}
	
	\begin{remark}
		Let $p^{m-1} \leq L < p^m$ and $L-1=\sum_{\alpha=1}^{m-1} d_{\alpha} p^{\alpha-1}+ d_m p^{m-1}$ for some $d_\alpha, d_m \in \mathbb{Z}_p$. Then either $d_m \neq 0$ or $d_m =0$. But $d_m=0$ means $d'_m=1, d'_\alpha=0, \forall \alpha=1,2,\dots,m-1$, where $L=\sum_{\alpha=1}^{m-1} d'_{\alpha} p^{\alpha-1}+ d'_m p^{m-1}$. This implies $d_\alpha=p-1, \forall \alpha=1,2, \dots, m-1$.
	\end{remark}
	
	\begin{note}
		If $L-1=\sum_{\alpha=1}^{m-1} d_{\alpha} p^{\alpha-1}+ d_m p^{m-1}$, then for $\beta p^{m-1} \leq i,j <(\beta+1) p^{m-1}$ (where $\beta \in \{0,1,2, \dots, d_m-1\}$), we have
		\begin{equation}
			\prod_{l=0}^{d_m-1} (i_m- l) =\prod_{l=0}^{d_m-1} (j_m- l)=0.
		\end{equation}
	\end{note}

	\begin{theorem}
		Let $p, m \in \mathbb{N}$ and $\pi$ be a permutation on the set $\{1,2, \dots, m-1\}$ with $\pi(1)=1$. Let $L \in \mathbb{N}$, where $ p^{m-1} \leq L < p^m $, $L-1=\sum_{\alpha=1}^{m-1} d_{\alpha} p^{\alpha-1}+ d_m p^{m-1}$ , $d_m, d_{\alpha} \in \mathbb{Z}_p, \forall \alpha$. Let $k$ be the smallest number such that $d_{\alpha} =0, \forall m-1 \geq \alpha \geq k$. If such a $k$ does not exist and $d_{\alpha} = (p-1)$ for all $1 \leq \alpha \leq m-1$, then we set $k=2$. Otherwise, we set $k=m$.  Let $f: \mathbb{Z}_p^m \rightarrow \mathbb{Z}_q$, where $q$ is any integer such that $p \mid q$ and the function $f$ is defined by
		\begin{equation}\label{eqn_main_f}
			\begin{split}
				f(x_1, x_2, \dots, x_m)= & \frac{q}{p} \sum_{\alpha=1}^{m-2} \!\!\! x_{\pi(\alpha)} x_{\pi(\alpha+1)} \\
				&+ g(x_1, \dots, x_{m-1})  \prod_{l=0}^{d_m-1} (x_m- l) \\
				&+ \sum_{\alpha=1}^{m} c_\alpha x_\alpha +c',
			\end{split}
		\end{equation}    
		where $ c' \in \mathbb{Z}_q$ and $g(x_1, x_2, \dots, x_{m-1})$ is any function $g : \mathbb{Z}_p^{m-1} \rightarrow \mathbb{Z}_q$. 
		Further, we define the functions $a^{\bm{\gamma}}$ for $\bm{\gamma}=\left(\gamma_1, \gamma_2, \dots, \gamma_{k} \right) \in \mathbb{Z}_p^{k}$, such that
		
		\begin{equation}
			\begin{split}
				a^{\bm{\gamma}}(x_1, x_2, \dots, x_m)= &f(x_1, x_2, \dots, x_m)+ \gamma_1 \frac{q}{p} x_{\pi(1)} \\
				&   + \frac{q}{p} \sum_{\alpha=2}^{k-1} \!\!  \gamma_\alpha x_{\alpha} +  \gamma_{k} \frac{q}{p} x_{m}.
			\end{split}
		\end{equation}
		Then, the sequence set 
		$\left\{\psi_L(a^{\bm{\gamma}}): \bm{\gamma} \in \mathbb{Z}_p^{k} \right\}$ is a $(q, p^{k}, L)$-GCS.
	\end{theorem}
	
	\begin{IEEEproof}
		We let $j= i+\tau$ and $\left(i_1, i_2, \dots, i_m\right), \left(j_1, j_2, \dots, j_m\right)$ be the $p$-ary representations of the natural numbers $i$ and $j$, respectively. As $\rho(\mathbf{a}) (-\tau)= \rho^* (\mathbf{a})(\tau)$, we shall only focus on the part where $\tau \geq 0$, in this proof.
		We divide this proof into two parts for easy understanding:
		\begin{mycases}
			
			\item Let $0 < \tau < d_m p^{m-1}$. In this case, we have two sub-cases:

			\begin{mysubcases}
				
				\item Let $i_{\pi(1)} \neq j_{\pi(1)}$. Then, we have
				
				\begin{equation}
					\begin{split}
						a^{\bm{\gamma}}(\vec{i})- 	a^{\bm{\gamma}}(\vec{j})= &\left(  f (\vec{i})- 	f(\vec{j}) \right) + \gamma_1 \frac{q}{p} \left(  i_{\pi(1)}  - j_{\pi(1)}\right)\\
						& + \frac{q}{p} \sum_{\alpha=2}^{k-1} \gamma_\alpha \left(  i_{\alpha} -  j_{\alpha}  \right) \\
						& + \gamma_{k} \frac{q}{p} \left(    i_{m} - j_{m}    \right).
					\end{split}
				\end{equation}
				This implies
				\begin{equation}
					\begin{split}
						\sum_{ \bm{\gamma} \in \mathbb{Z}_p^{k}}^{ }  \! \zeta_q^{a^{\bm{\gamma}}(\vec{i})- 	a^{\bm{\gamma}}(\vec{j})}=&  \zeta_q^{\left(  f (\vec{i})- 	f(\vec{j}) \right)} \!\!  \times \!\!  \sum_{ \gamma_{1}}^{ }  \zeta_q^{ \frac{q}{p} \gamma_{1} \left(    i_{\pi(1)} - j_{\pi(1)}    \right)}\\
						& \times \prod_{\alpha=2}^{k-1}  \sum_{ \gamma_\alpha}^{ } \zeta_q^{\frac{q}{p}  \gamma_\alpha \left(  i_{\alpha} - j_{\alpha} \right)}\\
						&\times  \sum_{ \gamma_{k}}^{ }  \zeta_q^{ \frac{q}{p} \gamma_{k} \left(    i_{m} - j_{m}    \right)}\\
						=& 0,
					\end{split}
				\end{equation}	
				as $\sum_{ \gamma_{1}=0}^{p-1}  \zeta_q^{ \frac{q}{p}  \gamma_{1} \left(    i_{\pi(1)} - j_{\pi(1)}    \right)}=0$.
				
				
				\item  Let $i_{\pi(1)} = j_{\pi(1)}$. We consider the two possibilities:
				\begin{enumerate}
					\item $i_m \neq j_m$. In this case, we would have
					$\sum_{ \gamma_{k}=0}^{p-1}  \zeta_q^{ \frac{q}{p}  \gamma_{k} \left(    i_{m} - j_{m}    \right)}=0$.
					
					\item $i_m = j_m$. In that case, $\exists$ an integer $v' \in \{1,2,\dots,m-1\}$ such that $i_{v'} \neq j_{v'}$. Procedure for this possibility is elaborated below.
				\end{enumerate}
				
				Without loss of generality, let us assume that $v$ is the greatest integer in $\{1, 2, \dots, m-1\}$ such that $i_{\pi(\alpha)}= j_{\pi(\alpha)}$ for $ \alpha \leq v$ and $\pi(v) \neq \beta$ for $\beta=1,2,\dots, k-1$. This implies that $i_{\pi(v+1)} \neq j_{\pi(v+1)}$. Here, we assume that two sets of integers $i^\delta$ and $j^\delta$ exists for $\delta=1, 2, \dots, p-1$ such that $i^\delta= \left(   i_1, i_2, \dots, (i_{\pi(v)}- \delta) \mod p, \dots, i_m\right)$, $j^\delta= \left(  j_1, j_2, \dots, (j_{\pi(v)}- \delta) \mod p, \dots, j_m\right)$. The existence of these sets implies that $i_m =j_m < d_m$. If such sets of integers do not exist, then the corresponding procedure is elaborated later. Let by $i''$ denote the $\left( i_1,  i_2, \dots, i_{m-1}\right)$, where $\vec{i}= \left(i'', i_m\right)$.\\			
				Now, we have
				\begin{equation}
					\begin{split}
						&a^{\bm{\gamma}}\left(\vec{i}\right)- 	a^{\bm{\gamma}}\left(\vec{i^\delta}\right)\\
						&= f \left(\vec{i}\right)- 	f\left(\vec{i^\delta}\right)+ \frac{q}{p}\sum_{\alpha=1}^{k-1} \left(i_\alpha-i^\delta_\alpha\right) \gamma_\alpha\\ 
						& \quad + \frac{q}{p} \left( i_m-i^\delta_m\right) \gamma_k \\
						& =\frac{q}{p} i_{\pi(v-1)} \left( \delta \mod p \right) + \frac{q}{p} i_{\pi(v+1)} \left( \delta \mod p \right) \\
						& \quad + \left(  g\left( \vec{i}'' \right)- g \left( \vec{i^\delta}'' \right)   \right)  \prod_{l=0}^{d_m-1} (i_m- l)\\
						& \quad + c_{\pi(v)} (\delta \mod p)\\
						&= \frac{q}{p} i_{\pi(v-1)} \left( \delta \mod p \right) + \frac{q}{p} i_{\pi(v+1)} \left( \delta \mod p \right)\\
						& \quad + c_{\pi(v)} (\delta \mod p)\\
					\end{split}
				\end{equation}
				as $\prod_{l=0}^{d_m-1} (i_m- l)=0$ in this case, and $i_\alpha= i^\delta_\alpha$ for $\alpha=1,2,\dots,k-1$ according to our assumption. Similarly, we have
				
				\begin{equation}
					\begin{split}
						&a^{\bm{\gamma}}\left(\vec{j}\right)- 	a^{\bm{\gamma}}\left(\vec{j^\delta}\right)\\
						&=\frac{q}{p} j_{\pi(v-1)} \left( \delta \mod p \right)  + \frac{q}{p} j_{\pi(v+1)} \left( \delta \mod p \right)\\
						& \quad + c_{\pi(v)} (\delta \mod p).\\
					\end{split}
				\end{equation}

				From the equations above, we have
				\begin{equation}
					\begin{split}
						&\left( a^{\bm{\gamma}}(\vec{i})- 	a^{\bm{\gamma}}(\vec{i^\delta}) \right)- 	\left( a^{\bm{\gamma}}(\vec{j})- 	a^{\bm{\gamma}}(\vec{j^\delta}) \right)\\
						&= \frac{q}{p} \left(i_{\pi(v+1)}-  j_{\pi(v+1)} \right) \left( \delta \mod p \right) \\
						\text{or}, & \left( a^{\bm{\gamma}}(\vec{i})- 	a^{\bm{\gamma}}(\vec{j})  \right)- 	\left( a^{\bm{\gamma}}(\vec{i^\delta})  - 	a^{\bm{\gamma}}(\vec{j^\delta}) \right)\\
						&= \frac{q}{p} \left(i_{\pi(v+1)}-  j_{\pi(v+1)} \right) \left( \delta \mod p \right), \\
					\end{split}
				\end{equation}
				which implies
				\begin{equation}
					\begin{split}
						&\zeta_q^{  \left( a^{\bm{\gamma}}(\vec{i})- 	a^{\bm{\gamma}}(\vec{j})  \right)- 	\left( a^{\bm{\gamma}}(\vec{i^\delta})  - 	a^{\bm{\gamma}}(\vec{j^\delta}) \right) }\\ 
						&= \zeta_p^{  \left(i_{\pi(v+1)}-  j_{\pi(v+1)} \right) \delta }\\
						&\text{or}, \sum_{\delta=1}^{p-1} \zeta_q^{  \left( a^{\bm{\gamma}}(\vec{i})- 	a^{\bm{\gamma}}(\vec{j})  \right)- 	\left( a^{\bm{\gamma}}(\vec{i^\delta})  - 	a^{\bm{\gamma}}(\vec{j^\delta}) \right)  } \\ 
						&= \sum_{\delta=1}^{p-1} \zeta_p^{  \left(i_{\pi(v+1)}-  j_{\pi(v+1)} \right) \delta}.\\
					\end{split}
				\end{equation}
				So, we have
				\begin{equation}
					\begin{split}
						& \text{or},  \sum_{\delta=1}^{p-1} \zeta_q^{ \left( a^{\bm{\gamma}}(\vec{i})- 	a^{\bm{\gamma}}(\vec{j})  \right)- 	\left( a^{\bm{\gamma}}(\vec{i^\delta})  - 	a^{\bm{\gamma}}(\vec{j^\delta}) \right) } = -1\\
						&	\text{or},  \sum_{\delta=1}^{p-1} \zeta_q^{ -\left( a^{\bm{\gamma}}(\vec{i^\delta})  - 	a^{\bm{\gamma}}(\vec{j^\delta}) \right)  }= -\zeta_q^{ - \left( a^{\bm{\gamma}}(\vec{i})- 	a^{\bm{\gamma}}(\vec{j})  \right) } \\
						&\text{or}, \zeta_q^{ \left( a^{\bm{\gamma}}(\vec{i})- 	a^{\bm{\gamma}}(\vec{j})  \right) } + \sum_{\delta=1}^{p-1} \zeta_q^{ \left( a^{\bm{\gamma}}(\vec{i^\delta})  - 	a^{\bm{\gamma}}(\vec{j^\delta}) \right)  } = 0.\\
					\end{split}
				\end{equation}
				
				The above procedure shows that we get the result if $i^\delta, j^\delta$ exists for $\delta= 1, 2, \dots, p-1$.

				Now, we consider the other possibility: Let such sets of $i^\delta, j^\delta$ do not exist, and on the other hand, we have $i_{\pi(1)} = j_{\pi(1)}$. As $0 < \tau < d_m p^{m-1}$, two situations may arise: either $i_{m} = j_{m}$ or $i_{m} \neq j_{m}$ . For $i_{m} \neq j_{m}$, we have

				\begin{equation}
					\sum_{ \gamma_{k}=0}^{p-1}  \zeta_q^{ \frac{q}{p}  \gamma_{k} \left(    i_{m} - j_{m}    \right)}=0.
				\end{equation}

				If $i_{m} = j_{m}$, that means $d_m p^{m-1} \leq i <j <d_m p^{m-1}+ p^{k-1}$ based on the assumption of the theorem. We claim that there should exists $\alpha_0 < k$ such that $i_{\alpha_0} \neq j_{\alpha_0}$. If the claim is false, then it implies $i_{\alpha} = j_{\alpha}, \forall \alpha = 1, 2, \dots, k-1$. But then, $d_m p^{m-1} \leq i <j <d_m p^{m-1}+ p^{k-1} \implies i_\alpha= j_\alpha=0$ for $\alpha=k,k+1, \dots,m-1$. This in turn says $i=j$ (contradiction). Hence,
				we have
				\begin{equation}
					\begin{split}
						&\sum_{ \bm{\gamma} \in \mathbb{Z}_p^{k}}^{ }  \zeta_q^{a^{\bm{\gamma}}(\vec{i})- 	a^{\bm{\gamma}}(\vec{j})}\\
						&=  \zeta_q^{\left(  f (\vec{i})- 	f(\vec{j}) \right)}  \times  \sum_{ \gamma_{1}}^{ }  \zeta_q^{ \frac{q}{p} \gamma_{1} \left(    i_{\pi(1)} - j_{\pi(1)}    \right)}\\
						& \quad \times \prod_{\alpha=2}^{k-1}  \sum_{ \gamma_\alpha}^{ } \zeta_q^{\frac{q}{p}  \gamma_\alpha \left(  i_{\alpha} - j_{\alpha} \right)} \!\! \times  \!\! \sum_{ \gamma_{k}}^{ }  \zeta_q^{ \frac{q}{p} \gamma_{k} \left(    i_{m} - j_{m}    \right)}\\
						&= 0,
					\end{split}
				\end{equation}
				as $\sum_{ \gamma_{\alpha_0}=0}^{p-1}  \zeta_q^{ \frac{q}{p}  \gamma_{\alpha_0} \left(    i_{\alpha_0} - j_{\alpha_0}    \right)}=0$.
				
			\end{mysubcases}
			
			
			\item Let $d_m p^{m-1} \leq \tau < L$. This automatically implies that $i_m \neq j_m$. So, we have
			
			\begin{equation}
				\begin{split}
					&\sum_{ \bm{\gamma} \in \mathbb{Z}_p^{k}}^{ }   \zeta_q^{a^{\bm{\gamma}}(\vec{i})- 	a^{\bm{\gamma}}(\vec{j})}\\
					&=   \zeta_q^{\left(  f (\vec{i})- 	f(\vec{j}) \right)} \times  \sum_{ \gamma_{1}}^{ }  \zeta_q^{ \frac{q}{p} \gamma_{1} \left(    i_{\pi(1)} - j_{\pi(1)}    \right)}\\
					&\quad \times \prod_{\alpha=2}^{k-1}  \sum_{ \gamma_\alpha}^{ } \zeta_q^{\frac{q}{p}  \gamma_\alpha \left(  i_{\alpha} - j_{\alpha} \right)} \times  \sum_{ \gamma_{k}}^{ }  \zeta_q^{ \frac{q}{p} \gamma_{k} \left(    i_{m} - j_{m}    \right)}\\
					&= 0,
				\end{split}
			\end{equation}	
			as $\sum_{ \gamma_{k}=0}^{p-1}  \zeta_q^{ \frac{q}{p}  \gamma_{k} \left(    i_{m} - j_{m}    \right)}=0$.
			
		\end{mycases}
		
		This proves the theorem.
	\end{IEEEproof}

	\begin{table}[ht]
		\centering
		$\left[   \phi_{19}  \left(   a^{\bm{\gamma}}\right)   \right]_{\bm{\gamma} \in \mathbb{Z}_4^2}= 
		\begin{+bmatrix}
			0     0     0     0     0     1     2     3     0     2     0     2   0     3     2     1     0     0     0\\
			
			0     1     2     3     0     2     0     2     0     3     2     1   0     0     0     0     0     1     2\\
			
			0     2     0     2     0     3     2     1     0     0     0     0   0     1     2     3     0     2     0\\
			
			0     3     2     1     0     0     0     0     0     1     2     3    0     2     0     2     0     3     2\\
			
			0     0     0     0     0     1     2     3     0     2     0     2    0     3     2     1     1     1     1\\
			
			0     1     2     3     0     2     0     2     0     3     2     1   0     0     0     0     1     2     3\\
			
			0     2     0     2     0     3     2     1     0     0     0     0   0     1     2     3     1     3     1\\
			
			0     3     2     1     0     0     0     0     0     1     2     3    0     2     0     2     1     0     3\\
			
			0     0     0     0     0     1     2     3     0     2     0     2  0     3     2     1     2     2     2\\
			
			0     1     2     3     0     2     0     2     0     3     2     1   0     0     0     0     2     3     0\\
			
			0     2     0     2     0     3     2     1     0     0     0     0    0     1     2     3     2     0     2\\
			
			0     3     2     1     0     0     0     0     0     1     2     3   0     2     0     2     2     1     0\\
			
			0     0     0     0     0     1     2     3     0     2     0     2   0     3     2     1     3     3     3\\
			
			0     1     2     3     0     2     0     2     0     3     2     1   0     0     0     0     3     0     1\\
			
			0     2     0     2     0     3     2     1     0     0     0     0    0     1     2     3     3     1     3\\
			
			0     3     2     1     0     0     0     0     0     1     2     3   0     2     0     2     3     2     1\\
		\end{+bmatrix}
		$
		\caption{$\mathbb{Z}_4$-valued GCS of length $19$, alphabet size $4$  and flock size $16$ constructed in \textit{Example \ref{eg_gcs}}}
		\label{table_gcs}
	\end{table}

	\begin{table*}
		\centering
		\begin{tabular}{|c|c|c|c|c|c|}
			\hline
			\textbf{Construction} & \textbf{Length}& \textbf{Flock Size}& \textbf{Alphabet Size} & \textbf{Constraints} & \textbf{Based On}\\
			\hline
			\hline
			\cite{paterson} & $2^m$ & $2^{k+1}$ & $q$ & $m \geq 1, k < m, 2 \mid q$ & GBF\\
			\hline
			\cite{ schmidt} & $2^m$ & $2^{k+1}$ & $q$ & $m \geq 1, k < m, 2 \mid q$ & GBF\\
			\hline
			\cite{ChenCCC2008}& $2^m$ & $2^k$ & $q$ & $m \geq 1, k \leq m, 2 \mid q$& GBF\\
			\hline
			\cite{ChenGCS-TIT} & $2^{m-1} + 2^v$& $4$& $q$ & $m \geq 2, v < m, 2 \mid q$ & GBF\\
			\hline
			\cite{ChenGCS2017} & $2^{m-1} + 2^v$ &  $2^{k+1}$ & $q$ &  $m \geq 2, v < m, 1 \leq k \leq m, 2 \mid q$ & GBF\\
			\hline
			\cite{ChenGCS2018} & $2^{m-1}+ \sum_{\alpha=1}^{k-1} a_\alpha 2^{\pi(m-k+ \alpha)+1}+ 2^v$& $2^{k+1}$& $q$& \makecell{$m \geq 2, v < m-k, k < m,$\\ $a_\alpha \in \{0,1\}, 2 \mid q$} & GBF\\
			\hline
			\cite{palashCS2021} & $2^m$& $2^{k+1}$& $q$ &  $m \geq 2, v < m, 1 \leq k <  m, 2 \mid q$ & GBF\\
			\hline
			\cite{ChenMOGCS2021}& $2^{m-1} + 2^v$ & $2^{k+1}$  & $q$ & $m \geq 2, v < m, 1 \leq k \leq m, 2 \mid q$   & GBF\\
			\hline
			\cite{ palashCCC} & $p_1^{m_1} p_2^{m_2} \cdots p_k^{m_k}$ & $p_1 p_2 \cdots p_k$ & $q$ & \makecell{$p_i$'s are prime, $m_i \geq 1, \forall i;$,\\  $k \geq 1$, $q \mid p_i, \forall i $ }& \makecell{Multivariable\\ Function}\\
			\hline
			\cite{ShenCCC} & $q^{m}$& $q^{k+1}$& $\lambda$ & $q \mid \lambda, k \leq m-1, m \geq 2, q \geq 2$ & EBF\\
			\hline
			Proposed & $L$ & $p^{k}$ & $q$ & \makecell{ $p, L \in \mathbb{N}, m \geq 1$, $ p^{m-1} \leq L < p^m $, \\ $L-1= \sum_{\alpha=1}^{m-1} d_{\alpha} p^{\alpha-1}$\\ $+ d_m p^{m-1}$,  \\ $d_m, d_{\alpha} \in \mathbb{Z}_p, \forall \alpha, p \mid q$, \\ $\pi$ is a permutation \\ on $\{1,2, \dots, m-1\}$, \\ with $\pi(1)=1$,\\ $k$ depends on $L$ } & EBF \\
			\hline
		\end{tabular}
		\caption{Comparison of Existing Direct Constructions with the Proposed One}
		\label{table_comparison1}
	\end{table*} 
	
	\begin{remark}
		The term $ g(x_1, \dots, x_{m-1})  \prod_{l=0}^{d_m-1} (x_m- l) $ in (\ref{eqn_main_f}) indicates that monomials of degree higher than $2$ can also be used to construct such GCSs.
	\end{remark}

	\begin{remark}\label{remark_GCS_p}
		For $L= p^{m-1}$ for some $m > 1$, $\overrightarrow{L-1}= \left(  p-1, \dots, p-1,0 \right) \in \mathbb{Z}_p^m$. In that case, $k= 2$. So, the corresponding function for the GCS would be given by
		\begin{equation}
			\begin{split}
				a^{\bm{\gamma}} (x_1, x_2, \dots, x_m)= &  \frac{q}{p} \sum_{\alpha=1}^{m-2} x_{\pi(\alpha)} x_{\pi(\alpha+1)} + \sum_{\alpha=1}^{m} c_\alpha x_{\alpha} \\
				&+ \gamma_{1} \frac{q}{p} x_{\pi(1)} + \gamma_{2} \frac{q}{p} x_m,
			\end{split}
		\end{equation}
		for $\bm{\gamma}= \left(  \gamma_{1}, \gamma_{2} \right) \in \mathbb{Z}_p^2$. Although, it seems that the set $\{ \psi_{p^{m-1}} (a^{\bm{\gamma}}) : \bm{\gamma} \in \mathbb{Z}_p^2 \}$ is a GCS of length $p^{m-1}$ and flock size $p^2$, it is not quite correct. The actual flock size is $p$, as $x_m=0$ for $\vec{i}= \vec{0}, \vec{1}, \dots, \overrightarrow{p^{m-1}-1}$. Hence, it produces a $(q,p, p^{m-1})$-GCS.
	\end{remark}

	Below, we validate the theorem with an example.
	
\begin{figure}
		\centering
		\includegraphics[height=5cm, width= 7cm]{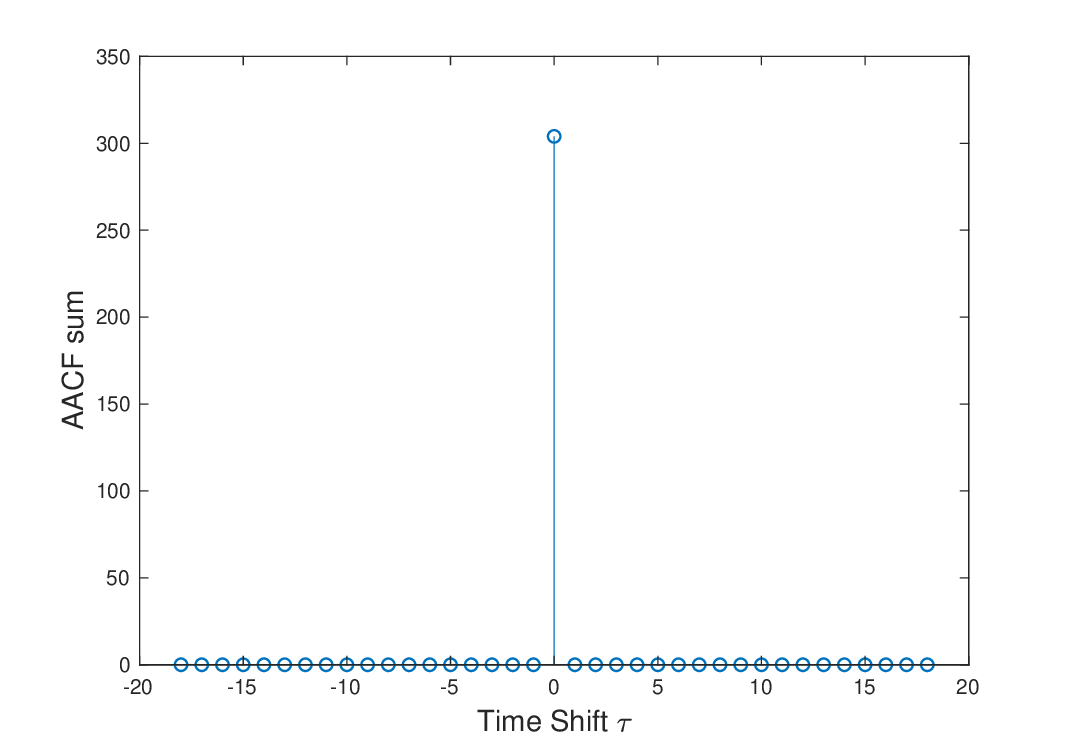}
		\caption{AACF plot for the GCS in \textit{Example \ref{eg_gcs}}}
		\label{aacf_gcs}
	\end{figure}

	\begin{example}\label{eg_gcs}
		Let $L= 19$ and $q=4$. We take the function $f: \mathbb{Z}_4^3 \rightarrow \mathbb{Z}_4$, which is defined by $f(x_1, x_2, x_3)= x_1 x_2 + 3 x_1 x_2 x_3+ 1$ with $\pi(1)=1, \pi(2)=2$, and satisfies the conditions of the theorem. The $4$-ary representation of the integer $L-1=19-1=18$ is given by $(2,0,1)$ and hence $k=2$. So, according to the theorem, we take the set of functions $a^{\bm{\gamma}}: \mathbb{Z}_4^3 \rightarrow \mathbb{Z}_4$, for $\bm{\gamma}= \left(   \gamma_1, \gamma_2 \right)$, which is defined by
		$a^{\bm{\gamma}}= f+ \gamma_1 x_1 + \gamma_2 x_3.$ 
		Then, $ \{ \psi_{19}  \left(   a^{\bm{\gamma}} \right) : \bm{\gamma} \in \mathbb{Z}_4^2 \} $ gives us a (4,16,19)-GCS, which is given in TABLE \ref{table_gcs}. The AACF sum of this GCS is plotted in Fig. \ref{aacf_gcs}.
	\end{example}

	\begin{remark}
		One of the constructions of GCSs in \cite{palashCCC} can be viewed as a special case of our construction when $p$ is a prime and $L=p^m$. This can be easily seen from \textit{Remark \ref{remark_GCS_p}}. Moreover, the proposed construction can produce arbitrary lengths and alphabet sizes, while in \cite{palashCCC}, the constructed GCS is restricted by the alphabet size.
	\end{remark}
	
	\begin{remark}
		As the proposed construction has a flock size of the form $p^k$ for $k \geq 1$, by using \textit{Lemma \ref{lemma_PMEPR}}, we can say that the PMEPR of each sequence is upper bounded by $p^k$, for some $p \in \mathbb{N}$ used in the construction.
	\end{remark}

	\subsection{Comparison with the State-of-the-Art}
	In this subsection, we compare the proposed construction with the existing function-based constructions with respect to their parameter constraints. The detailed comparison is given in TABLE \ref{table_comparison1}.

	\section{Conclusion}
	
	In this paper, we have proposed a new direct construction of GCS of arbitrary alphabet size and length with the help of EBFs. As the title suggests, the proposed construction achieves the maximum possible flexibility within restrictions of only one parameter, which is flock size. The proposed GCSs show bounded and low PMEPRs for some cases. Also, in this work, we are considering higher-order EBFs to produce GCSs.

	\balance

	\bibliographystyle{IEEEtran}
	\bibliography{Abhishek_GCS_ref}

\end{document}